\documentclass[12pt]{article}
\usepackage{graphics,amssymb,latexsym,amsmath,epsfig,color,graphicx}

\newcommand\pubnumber{}
\newcommand\pubdate{\today}

\def\triumf{TRIUMF, 4004 Wesbrook Mall Vancouver, BC V6T 2A3, Canada}

\def\Title#1{\begin{center} {\Large #1 } \end{center}}
\def\Author#1{\begin{center}{ \sc #1} \end{center}}
\def\Address#1{\begin{center}{ \it #1} \end{center}}

\newcommand\pubblock{\rightline{\begin{tabular}{l} \pubnumber\\
         \pubdate  \end{tabular}}}
\newenvironment{Abstract}{\begin{quotation}  }{\end{quotation}}
\newenvironment{Presented}{\begin{quotation} \begin{center} 
             PRESENTED AT\end{center}\bigskip 
      \begin{center}\begin{large}}{\end{large}\end{center} \end{quotation}}
\def\Acknowledgements{\bigskip  \bigskip \begin{center} \begin{large}
             \bf ACKNOWLEDGEMENTS \end{large}\end{center}}

\def\beq{\begin{equation}}
\def\eeq#1{\label{#1}\end{equation}}
\def\eeqn{\end{equation}}

\def\beqa{\begin{eqnarray}}
\def\eeqa#1{\label{#1}\end{eqnarray}}
\def\eeqan{\end{eqnarray}}

\let\bar=\overbar

\def\Dslash{\not{\hbox{\kern-4pt $D$}}}
\def\dslash{\not{\hbox{\kern-2pt $\del$}}}

\def\msb{{\bar{\ssstyle M \kern -1pt S}}}

\begin{document}
\begin{titlepage}
\pubblock

\vfill
\Title{Lattice results on charmonium}
\vfill
\Author{ Daniel Mohler}
\Address{\triumf}
\vfill
\begin{Abstract}

\end{Abstract}
\vfill
\begin{Presented}
The 5th International Workshop on Charm Physics\\ (Charm 2012).
\end{Presented}
\vfill
\end{titlepage}
\def\thefootnote{\fnsymbol{footnote}}
\setcounter{footnote}{0}

\section{Introduction}

Lattice QCD (LQCD) is a tool for {\it{ab initio}} calculations of the theory of quantum chromodynamics (QCD) in the non-perturbative regime. As such it is ideally suited for calculations of the hadron spectrum and to calculate properties of hadrons. In the charmonium spectrum, many observed states above multiparticle threshold do not fit into the picture resulting from simple potential models such as \cite{Godfrey:1985xj}. While these states can be studied with a variety of models, QCD calculations are very desirable. Furthermore, the rich spectrum observed in charmonium is an ideal benchmark for lattice calculations. The strange and light quark masses and the lattice scale are usually determined from light-quark observables. This makes calculations of the charmonium spectrum an ideal benchmark calculation.

While calculations in LQCD were limited to larger than physical pion masses for many years, this obstacle has been overcome in recent years and there have been first calculations at \cite{Aoki:2009ix,Durr:2010vn} or close to \cite{Aoki:2008sm} physical pion masses using at least 2 flavors of dynamical quarks.

As an example, some ground state masses of light-quark mesons and baryons have been calculated with full control of systematic uncertainties \cite{Durr:2008zz}. In these proceedings I will highlight corresponding progress for the calculation of the charmonium spectrum from Lattice QCD. In Section \ref{methodology} some basic issues for lattice calculations are briefly reviewed. Apart from discussing continuum, infinite volume and chiral extrapolations a well-established method to extract excited states on the lattice is presented. In section \ref{recentsim} results from various recent simulations are collected. Section \ref{conclusion} provides some brief concluding remarks.

\section{\label{methodology}Lattice methodology}

\subsection{\label{extrapolations}A threefold of extrapolations}

In LQCD observables are calculated using Monte-Carlo methods. Any such simulation on a finite number of configurations, in our case the finite number of dynamic gauge configurations within an ensemble, will have a stochastic uncertainty associated with the estimate. In addition to the stochastic uncertainty, rigorous quantitative comparisons with experimental results can only be made after systematic extrapolation to the continuum and infinite volume limits and after extrapolation or interpolation to physical quark masses. How important these uncertainties are depends on the observable and on the lattice action used. In this section, we discuss the role of these extrapolations and provide a short explanation how these issues are handled in practice.

While a fully systematic treatment and a quantitative comparison to experiment mandate these extrapolations, many qualitative insights can be obtained from simulations where some of the uncertainties associated with continuum, infinite volume or chiral extrapolations have not been fully controlled. However, caution has to be exercised when interpreting the resulting data. In Section \ref{recentsim} we will highlight this using the example of the charmonium hyperfine splitting, which can be quite sensitive to discretization effects.

\subsubsection{Continuum limit}

Observables measured on the lattice contain discretization errors which depend on the lattice spacing $a$. To compare to experiment an extrapolation to the continuum has to be performed, which corresponds to the limit $a(\beta,m_{q_i})\rightarrow 0$, with the inverse gauge coupling $\beta=\frac{6}{g^2}$ and the quark masses $m_{q_i}$. Given multiple lattices of the same physical size with different lattice spacings one can fit the data with the functional form expected for the leading discretization errors and obtain values in the continuum limit. In practice this is a delicate business in simulations with dynamical fermions. In particular the extrapolation should be performed along lines of constant physics and the physical volume is usually not constant for the initial estimate of simulation parameters. It should be stressed, that while complicated in detail, this is a straight-forward procedure for many observables.

\subsubsection{Infinite volume limit}

For a given volume, interactions due to the periodicity of the lattice lead to finite volume effects. Provided a large enough lattice, these cause exponentially small corrections to the spectrum of hadrons \cite{Luscher:1985dn}. These are usually given by the lightest particle in the spectrum, the pion, and are therefore of order $\mathrm{exp}(-m_\pi L)$, where $m_\pi$ is the pion mass and $L$ is the size of the box. It is often assumed that these effects can be neglected for lattices where the dimensionless quantity $m_\pi L> 4$. A better approach consists of measuring all observables on boxes of increasing size (keeping all other simulation parameters constant) and performing an extrapolation to the infinite volume limit. 

\subsubsection{Physical quark masses}

While calculation at physical quark masses are now possible, present simulations still commonly employ larger than physical light-quark masses. In this case an extrapolation to physical quark masses has to be performed. This is often accomplished by taking a fit form motivated by chiral effective field theories (for example Chiral Perturbation theory ($\chi$PT) in the case of light mesons or Heavy Meson Chiral Perturbation Theory in the case of heavy-light mesons). In the lattice literature this extrapolation is often referred to as ``chiral extrapolation'', even in cases where the actual extrapolation is to non-vanishing light-quark masses.

\subsection{\label{varmethod}Excited states and the variational method}

On the lattice correlators in Euclidean space-time are measured. It can be shown \cite{Gattringer:2010zz} that these behave as
\begin{align}
\left <\hat{O}_2(t)\hat{O}_1(0)\right >_T &\propto\sum_{n}e^{-tE_n}<0|\hat{O}_2|n><n|\hat{O}_1|0>\; .
\end{align}
The left hand side can be expressed as a path integral and its discretization can be used to calculate this expression using Monte-Carlo techniques. In the simplest example, interpolating field operators $\hat{O}_1$ and $\hat{O}_2$ that create and destroy a hadron with specific flavor and quantum numbers can be used to obtain information about the hadron spectrum. As an example, the interpolator $\bar{d}\gamma_5 u$ could be used for a $\pi^+$.

Using interpolators of definite quantum numbers it is therefore straight-forward to calculate the ground state in each channel, which can be obtained from the large time behavior of the Euclidean correlator. To extract not only the ground states but the low-lying spectrum one commonly calculates a matrix of correlators at some source time slice $t_i$ and for every sink time slice $t_f$ 
\begin{align}
C_{ij}(t=t_f-t_i)&=\sum_{t_i}\langle 0|O_i(t_f)O_j^\dagger(t_i)|0\rangle\\
&=\sum_n\mathrm{e}^{-tE_n} \langle 0|O_i|n\rangle \langle n|O_j^\dagger|0\rangle\;.\nonumber
\end{align}
For charmonium interpolators of definite $J^{PC}$, where $J$ is spin, $P$ is parity and $C$ is charge conjugation would be used\footnote{In practice this is complicated by the fact that, due to the loss of rotational symmetry, an infinite tower of continuum states contributes to a given irreducible representation.}. The low-lying spectrum is extracted by solving the generalized eigenvalue problem
\begin{align}
C(t)\vec{\psi}^{(k)}&=\lambda^{(k)}(t)C(t_0)\vec{\psi}^{(k)}\;,\\
\lambda^{(k)}(t)&\propto\mathrm{e}^{-tE_k}\left(1+\mathcal{O}\left(\mathrm{e}^{-t\Delta E_k}\right)\right)\;.\nonumber
\end{align}
for each time slice. At large time separation only a single state contributes to each eigenvalue. This procedure is known as the variational method \cite{Luscher:1990ck,Michael:1985ne,Blossier:2009kd}. It is used for several of the results presented in Section \ref{recentsim}.
\newpage
\section{\label{recentsim}Recent charmonium simulations}

\subsection{\label{fnalmilc}Results from FNAL/MILC}

\begin{table}[htb]
\begin{center}
\begin{tabular}{|l|l|l|l|l|l|}
  \hline
  Study  & sea & Charm &  Ops & States & Comment \\
  \hline
  2009 FNAL/MILC & $2+1$    &  FNAL & 2    & low S, P & $a \ge 0.09$ fm \\
  2012 FNAL/MILC & $2+1$    &  FNAL & many & many     & $a \ge 0.045$ fm \\
  Current FNAL/MILC & $2+1+1$ &  HISQ &  2   & low S, P & $a \ge 0.06$ fm \\
  Future FNAL/MILC & $2+1+1$  &  HISQ & many & many     & $a \ge 0.06$ fm \\
  \hline
\end{tabular}
\caption{\label{fnal_milc_cbarc}Analysis campaigns within the Fermilab Lattice and MILC collaboration charmonium project. 2009 results refer to \cite{Burch:2009az}. In addition to the analysis campaigns using lattices with  2+1 dynamical flavors of asqtad quarks, there are also current and future campaigns using gauge configurations with 2+1+1 flavors of highly improved staggered quarks (HISQ), where the charm quark is also included in the sea.}
\end{center}
\end{table}

\begin{figure}[tbh]
\centering
\includegraphics[height=70mm,clip]{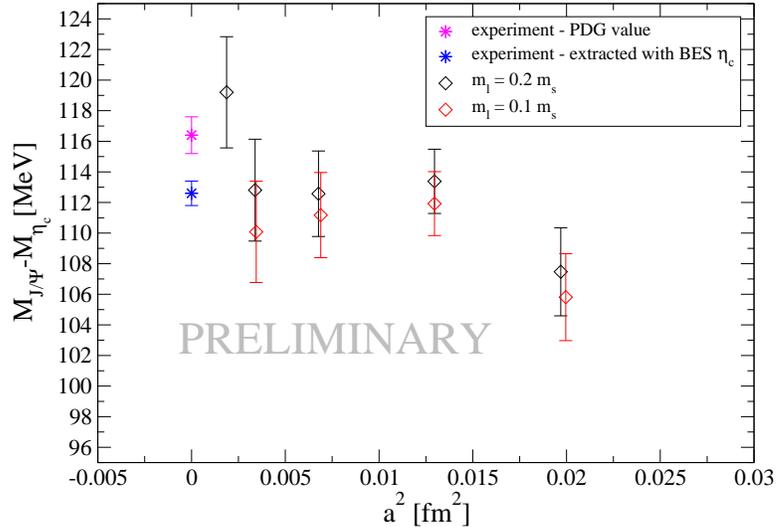}
\caption{Preliminary results for the hyperfine splitting $M_{J/\Psi}-M_{\eta_c}$ as a function of the lattice spacing squared. For further details please refer to the text.}
\label{fig:hf_FNALMILC}
\end{figure}

In this section some preliminary results obtained by the Fermilab Lattice and MILC collaborations are discussed. The goal of their charmonium project is a precision study of the low-lying charmonium states also including higher spin states including a full error budget for uncertainties stemming from the Monte Carlo analysis and for systematic uncertainties. Table \ref{fnal_milc_cbarc} provides an overview of past and current analysis campaigns related to charmonium. Results presented in this publication are preliminary results from the 2012 analysis campaign using 2+1 flavor asqtad configurations generated by the MILC collaboration \cite{Aubin:2004wf,Bernard:2001av}. 

Figure \ref{fig:hf_FNALMILC} shows the 1S-hyperfine splitting $M_{J/\Psi}-M_{\eta_c}$ as a function of the lattice spacing squared compared to experiment. Five different lattice spacings ranging from a rather coarse $a\approx0.15\mathrm{fm}$ to a very fine $a\approx0.045\mathrm{fm}$ are used for two different masses of light sea quarks $m_l$, specified as a fraction of the strange quark mass $m_s$. Note that both light quark masses are heavier than physical, necessitating an extrapolation. Experiment values from the PDG \cite{Beringer:1900zz} and a recent result by the BES collaboration \cite{BESIII:2011ab} are indicated by magenta and blue stars respectively. While the plot does not yet show a continuum or chiral extrapolation, which will be performed on the final dataset, the data is consistent with recent experimental determinations. Within the renormalization scheme used, there is a small but significant sea quark mass dependence of the results. From Figure \ref{fig:hf_FNALMILC} it is also evident, that the use of an improved heavy-quark action, the so-called Fermilab action \cite{ElKhadra:1996mp} leads to rather mild discretization effects. While the current results are fully consistent with experiment we would like to point out that contributions from quark annihilation have not yet been included and that the physical $\eta_c$ has a non-negligible width, which is neglected in this simulation. The effect of the quark-annihilation contributions will be discussed below.

\begin{figure}[htb]
\centering
\includegraphics[height=70mm,clip]{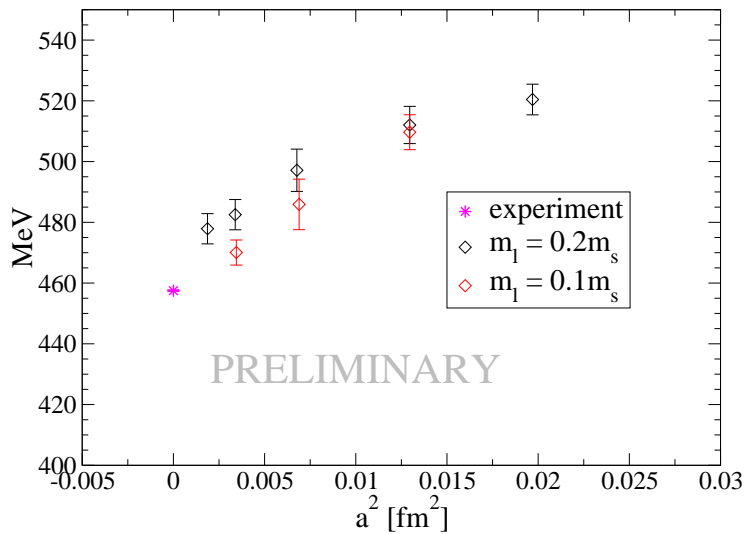}
\caption{Preliminary results for the $\overline{1S}$ - $\overline{1P}$ splitting as a function of the lattice spacing squared. For further details please refer to the text}
\label{fig:1s1p_FNALMILC}
\end{figure}

In addition to spin-dependent quantities like the hyperfine splitting, one can also take a look at spin-independent splittings. As an example one can define mass centroids for the $1S$ and $1P$ states
\begin{align}
M_{\overline{1S}}&=\frac{1}{4}(M_{\eta_c}+3M_{J/\Psi})\; ,\\
M_{\overline{1P}}&=\frac{1}{9}(M_{\chi_{c0}}+3M_{\chi_{c1}}+5M_{\chi_{c2}})\; .\nonumber
\end{align} 

Figure \ref{fig:1s1p_FNALMILC} shows the resulting $\overline{1S}$ - $\overline{1P}$ splitting as a function of the lattice spacing squared. The lattice data for two different sea quark masses has been plotted as black and red diamonds and the experimental data from the PDG is indicated by a magenta star. Again, within the renormalization scheme used the data show a non-negligible sea-quark mass dependence at fine lattice spacings. While extrapolations have not yet been performed the results are consistent with the experimental value.

\subsection{\label{HPQCD}Charmonium hyperfine splitting from HPQCD}

\begin{figure}[htb]
\centering
\includegraphics[height=70mm,clip]{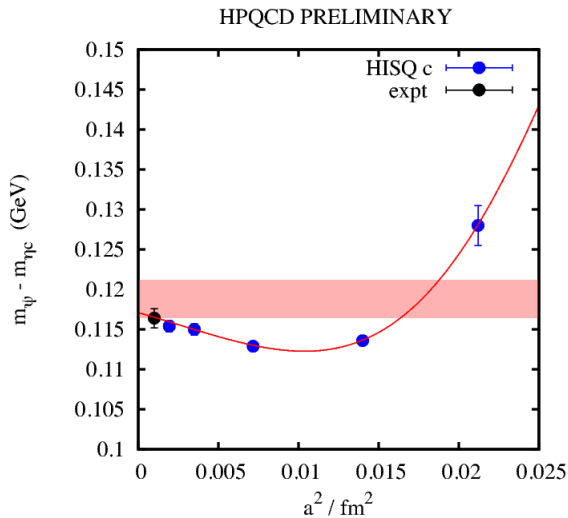}
\caption{Results for the hyperfine splitting from HISQ charm quarks by the HPQCD collaboration. Plot taken from \cite{Follana:2010zz}. For further details please refer to the text.}
\label{fig:hf_HPQCD}
\end{figure}

Figure \ref{fig:hf_HPQCD} shows results from the HPQCD-collaboration \cite{Follana:2010zz} for the charmonium hyperfine splitting \footnote{In the meantime HPQCD has updated their calculation and the results can now be found in \cite{Donald:2012ga}.}. For the simulation highly improved staggered quarks (HISQ) are used for the valence charm quarks and the sea contains 2+1 light flavors of asqtad staggered quarks. The simulation data is displayed as blue circles, along with a continuum extrapolation of the results (red line). The final results are indicated by the red band and the numerical value obtained is $118.8\pm2.4$MeV. For comparison the PDG value for the hyperfine splitting is displayed as a black circle.

\subsection{\label{disconnected}Disconnected contributions to the hyperfine splitting}

\begin{figure}[htb]
\centering
\includegraphics[clip, height=1.2cm]{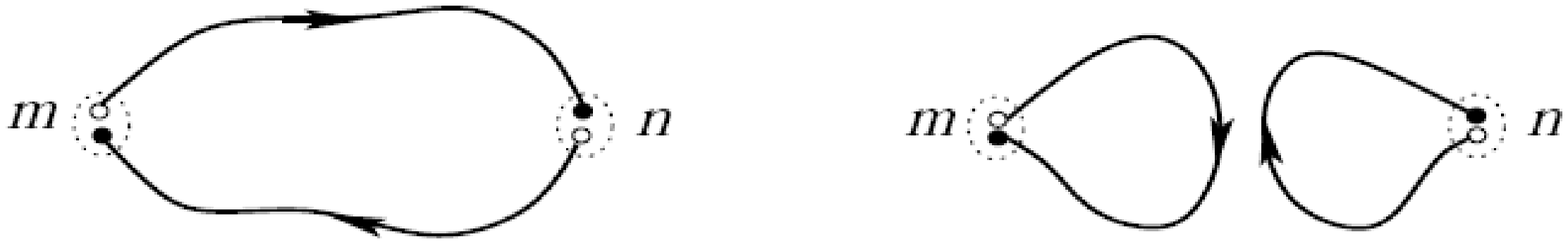}\\
\caption{Illustration of quark-line diagrams contributing to Charmonium correlators. The calculation of contributions from the  diagram on the right is technically demanding and these contributions should be small. Therefore this contribution is often omitted.}
\label{fig:disconnected}
\end{figure}

Before presenting further results, it is worthwhile to briefly discuss the role of disconnected contributions to the hyperfine splitting. Their determination is extremely challenging and therefore these contributions are often neglected or estimated from theory. For $\bar{q}q$ mesons the relevant quark-line diagrams are displayed in Figure \ref{fig:disconnected}. Neglecting annihilation effects corresponds to omitting the quark line diagram displayed on the right side of the figure.

For the results of the hyperfine splitting from HPQCD presented in the previous subsection, contributions from disconnected diagrams are estimated from perturbation theory which leads to a shift of simulation results enlarging the hyperfine splitting. Correspondingly the central value red band in Figure \ref{fig:hf_HPQCD} with the final results indicates a hyperfine-splitting larger than the continuum extrapolated value. Another strategy consist of estimating the effects of annihilation diagrams by direct calculation. This approach has been taken by the MILC collaboration in \cite{Levkova:2010ft}. The authors find that disconnected contributions to the hyperfine splitting \emph{reduce} the hyperfine splitting by 1-4MeV. It should be emphasized that novel methods \cite{Peardon:2009gh,Liu:2012ze} should lead to a more precise determination of annihilation effects in future direct simulations.

\subsection{\label{expolratory}Two exploratory studies}

In this section we take a look at results from two recent exploratory studies \cite{Mohler:2011ke,Mohler:2012na}. While their focus is on heavy-light mesons, the low-lying charmonium spectrum has been calculated as well. In contrast to the results discussed in the previous subsection the aim of these studies is to push the boundaries with regard to extracting excited states and to move towards calculations of hadron resonances on the lattice. As these are more difficult problems, current simulations commonly do not include a full treatment of systematic uncertainties. In particular they often lack continuum and infinite volume extrapolations.

\begin{table}[tbh]
\begin{center}
\begin{tabular}{ccccc}
\hline
$N_L^3\times N_T$ & $a$[fm] & $L$[fm] & \#configs & $m_\pi$[MeV]\\ 
\hline
$32^3\times32$ & 0.0907(13) & 2.9 & $\ge198$ & $156\le m_\pi\le702$ \\
\hline
\end{tabular}
\caption{\label{confs_pacs-cs}Parameters of the PACS-CS lattices \cite{Aoki:2008sm} used in \cite{Mohler:2011ke}. $N_L$ and $N_T$ denote the number of lattice sites in spatial and time directions, $a$ denotes the lattice spacing and $L$ denotes the physical volume. For the results presented in these proceedings the values from the ensemble with the lightest pion mass $M_\Pi\approx156$MeV are used. Please refer to \cite{Mohler:2011ke} for further details about the simulation.}
\end{center}
\end{table}

The first study \cite{Mohler:2011ke} uses $N_f=2+1$ Wilson-Clover configurations generated by the PACS-CS collaboration \cite{Aoki:2008sm}. For charm , the Fermilab prescription \cite{ElKhadra:1996mp,Bernard:2010fr} is used. In the following, results from the lightest pion mass, which is close to the physical pion mass will be presented. For details about the gauge configurations please refer to Table \ref{confs_pacs-cs}.

\begin{table}[tbh]
\begin{center}
\begin{tabular}{ccccc}
\hline
$N_L^3\times N_T$ & $a$[fm] & $L$[fm] & \#configs & $m_\pi$[MeV]\\ 
\hline
$16^3\times32$ & 0.1239(13) & 1.98 & 280 & 266(3)(3) \\
\hline
\end{tabular}
\caption{\label{conf_anna}Parameters for the Clover-Wilson lattices used in \cite{Mohler:2012na}. For an explanation of the parameters please refer to Table \ref{confs_pacs-cs}.}
\end{center}
\end{table}

The second study uses configurations with $N_f=2$ flavors of nHYP-smeared \cite{Hasenfratz:2007rf} Wilson-Clover quarks generated for reweighting studies \cite{Hasenfratz:2008fg,Hasenfratz:2008ce}. For the charm quarks the Fermilab method \cite{ElKhadra:1996mp,Bernard:2010fr} is employed. Details about the gauge configurations are listed in Table \ref{conf_anna}. Further information about this simulation can be found in \cite{Mohler:2012na}.

\begin{figure}[htb]
\centering
\includegraphics[height=70mm,clip]{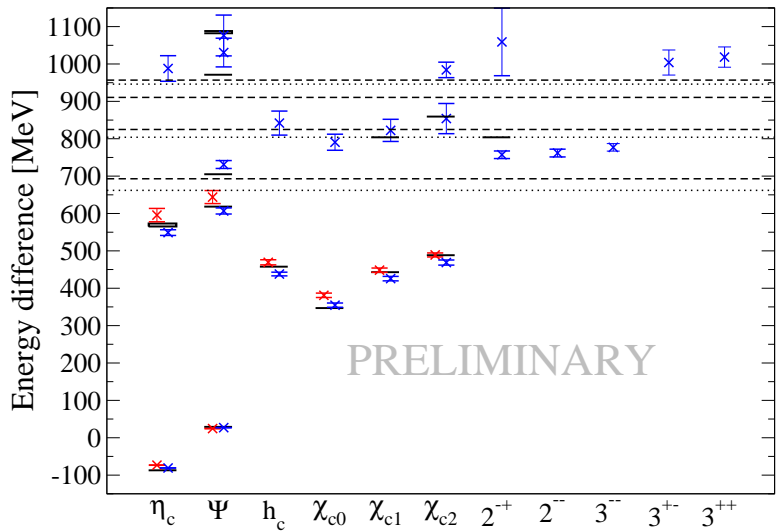}
\caption{Charmonium spectrum as extracted from the lattice from the two different studies described in detail in the text. The red crosses (displaces slightly to the left) are from \cite{Mohler:2011ke} while the blue crosses (displaced slightly to the right) display results from \cite{Mohler:2012na}. The error bars include statistic errors and the systematic uncertainty from the scale-setting procedure. Experimentally observed states are plotted as black bars or filled boxes. The level corresponding to the well-established X(3872) has been plotted for both choices of possible quantum numbers ($1^++$ or $2^{-+}$).}
\label{fig:spectrum_charm_DM}
\end{figure}

\begin{table}[bht]
\begin{center}
\begin{tabular}{|c|c|c|}
\hline
Mass difference & This study [MeV] & Experiment [MeV]\\
\hline
\hline
$\overline{1P}-\overline{1S}$ & $441.7\pm4.0\pm4.6$ & $457.5\pm0.3$\\
\hline
$\overline{2S}-\overline{1S}$ & $592.3\pm4.9\pm6.2$ &  $606.1\pm1.0$\\
\hline
1S hyperfine & $107.9\pm0.3\pm1.1\pm_0^{2.2}$ & $116.6\pm1.2$\\
\hline
1P spin-orbit & $39.7\pm2.1\pm0.4$ &  $46.6\pm0.1$ \\
\hline
1P tensor & $11.02\pm0.87\pm0.12$ &  $16.25\pm0.07$ \\
\hline
1P hyperfine & $3.7\pm2.7$& $-0.10\pm0.22$ \\
\hline
2S hyperfine & $57.9\pm2.0$ & $49\pm4$ \\
\hline
2P spin-orbit & $24.6\pm15.7\pm0.3$ & - \\
\hline
2P tensor & $2.2\pm4.3$ & - \\
\hline
$\overline{2P}-\overline{1S}$ & $836.4\pm30.5\pm8.8$ & - \\
\hline
$2M_{\overline{D_s}}-M_{\overline{\bar{c}c}}$ & $1065.6$ & $1084.8\pm0.8$ \\
\hline
\end{tabular}
\end{center}
\caption{\label{table:anna}Mass differences in the charmonium spectrum in MeV compared to experimental values (calculated from \cite{Beringer:1900zz}; the value for the 1P hyperfine splitting is from \cite{Olsen:2012xn}). Bars denote spin-averaged values. For the results of this paper, the first error denotes the statistical uncertainty and the second error denotes the uncertainty from setting the lattice scale. It is stressed that the gauge ensembles used do not allow for a continuum and infinite volume extrapolation. Consequently qualitative but not quantitative agreement is expected. In the last line we also provide the splitting $2M_{\overline{D_s}}-M_{\overline{c\bar{c}}}$ which can be directly compared to the results quoted by the Fermilab lattice and MILC collaborations \cite{Burch:2009az} and also to the value of $2M_{D_s}-M_{\eta_c}$ quoted by HPQCD in \cite{Follana:2006rc}.}
\end{table}

Figure \ref{fig:spectrum_charm_DM} shows the difference between several low-lying charmonium masses and the mass of the spin-averaged $1S$ ground state $M_{\overline{1S}}=\frac{1}{4}(M_{\eta_c}+3M_{J/\Psi})$. The quantum number channels are denoted by either their particle names (for the low-lying states) or by their $J^{PC}$, where $J$ is the Spin, $P$ stands for the parity quantum number and $C$ is charge conjugation. For both studies, there is good qualitative agreement with the low-lying spectrum observed in experiment and the $1S$, $1P$ and $2S$ states are well reproduced. In addition to the low-lying states extracted in both simulations, the full sets of $1D$ and $2P$ states are extracted in the more recent study \cite{Mohler:2012na}. The high statistical precision needed for this has been achieved through the \emph{distillation} technique proposed in \cite{Peardon:2009gh}. To illustrate the good statistical precision, Table \ref{table:anna} summarizes the charmonium results. For a full discussion of these results please refer to \cite{Mohler:2012na}.

\subsection{\label{hsc}Excited states from anisotropic Wilson-Clover lattices}

The Hadron Spectrum Collaboration has recently studied \cite{Liu:2012ze} ground and excited charmonium states using anisotropic Wilson-Clover lattices with volumes $16^3\times 128$ and $24^3\times 128$. The spatial lattice spacing is $a_s\approx0.12$fm and the anisotropy between the lattice spacing in spatial and time directions is $\xi=3.5$. The sea-quark pion mas for their simulations is $m_\pi=396$MeV. They use the variational method with a basis containing interpolating fields with up to three covariant derivatives. A similar basis has previously been used for studies of light-quark mesons \cite{Dudek:2010wm}. Before turning to results let us discuss two notable features about this simulation. First, the distillation technique \cite{Peardon:2009gh} allows calculations with small statistical uncertainty, even for some of the higher excitations. In addition advanced spin identification techniques \cite{Dudek:2010wm} allows for unambiguous assignment of certain lattice states to a particular continuum spin. In particular, this enables the authors of \cite{Liu:2012ze} to clearly identify Spin 4 states and states with spin-exotic quantum numbers.

\begin{figure}[htb]
\centering
\includegraphics[height=70mm,clip]{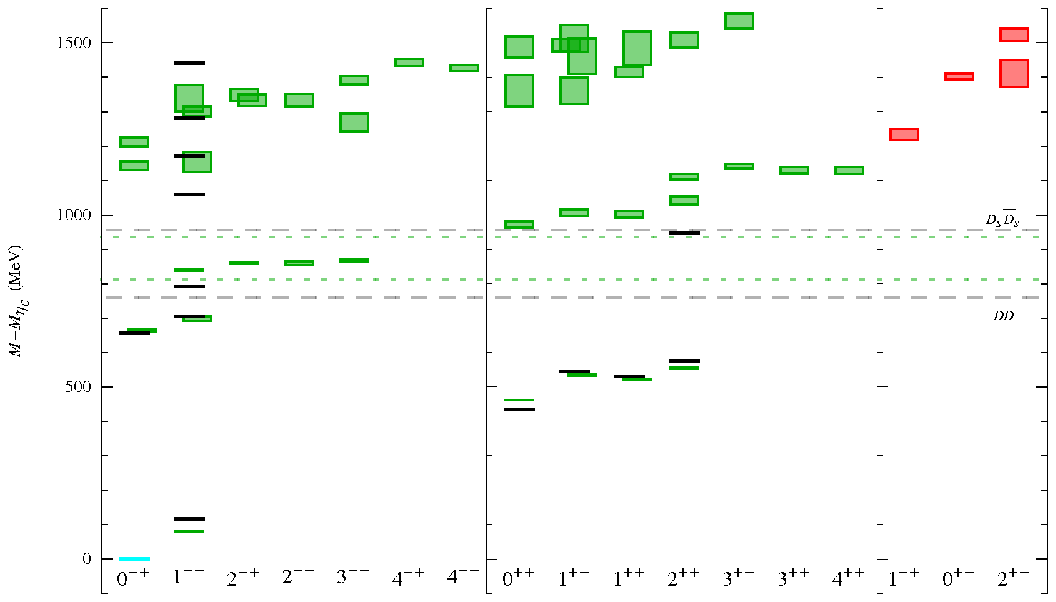}
\caption{Difference between the mass of charmonium states and the $\eta_c$ mass $M_{\eta_c}$. A large number of states including spin-exotic states has been identified. Figure taken from \cite{Liu:2012ze}}
\label{fig:spectrum_charm_HSC}
\end{figure}

Figure \ref{fig:spectrum_charm_HSC} shows the results for the charmonium spectrum from \cite{Liu:2012ze}. A large number of conventional and non-conventional states has been extracted. The states displayed as red boxes on the right side of the figure correspond to spin-exotic states which can be cleanly identified by means of spin-identification \cite{Dudek:2010wm}. The authors point out that some of the non spin-exotic states fit into a possible multiplet of hybrid states. For higher-lying states the authors are careful to point out that these states will be influenced by unphysical thresholds and should only coincide with physical states up to the hadronic width of these states. Notice that the distillation technique \cite{Peardon:2009gh} allows for the inclusion of these thresholds in a straight-forward manner. For more information please refer to \cite{Liu:2012ze}.

To assess the discretization effects present in this study it is instructive to take a look at the $1S$ hyperfine splitting $M_{HFS}=80\pm 1$ MeV. Unlike the results from the MILC and HPQCD collaborations this result differs significantly from the experimental value. The difference illustrates the typical size of discretization effects. One has to stress that this can be improved by the use of an improved heavy quark action or by a continuum extrapolation. We will return to this in subsection \ref{BMWQCDSF} where it is shown that these issues can be overcome by a proper continuum extrapolation.

\subsection{\label{BMWQCDSF}Results from the BMW and QCDSF collaborations}

\begin{table}[tbh]
\begin{center}
\begin{tabular}{ccccc}
\hline
$N_L^3\times N_T$ & $a$[fm] & $L$[fm] & $m_\pi$[MeV]\\ 
\hline
$24^3\times48/ 32^3\times 64$ & 0.0795(3) & 1.9/ 2.5 & 442 \\
$24^3\times48/ 32^3\times 64$ & 0.0795(3) & 1.9/ 2.5 & 348 \\
\hline
\end{tabular}
\caption{\label{qcdsf_slinc}Parameters of the QCDSF lattices used in \cite{Bali:2011dc}}
\end{center}
\end{table}

In a joint project, the Budapest-Marseille-Wuppertal(BMW) and QCDSF collaborations are investigating the charmonium spectrum using configurations with $2+1$ flavors of sea quarks.
To this end they use both Wilson-Clover gauge configurations generated by BMW \cite{Durr:2010aw} and SLiNC gauge configurations  \cite{Cundy:2009yy} generated by QCDSF. The BMW configurations span a wide range of lattice spacings $0.054\mathrm{fm}\le a\le0.092\mathrm{fm}$ and pion masses $120\mathrm{MeV}\le M_\pi\le 520\mathrm{MeV}$ on volumes of size $L^3\times T=32^3\times 64$ and  $64^3\times144$. For the QCDSF configurations the parameters of used configurations are listed in Table \ref{qcdsf_slinc}. For both actions the respective fermion action is also used for the charm quarks. 

\begin{figure}[htb]
\centering
\includegraphics[height=70mm,clip]{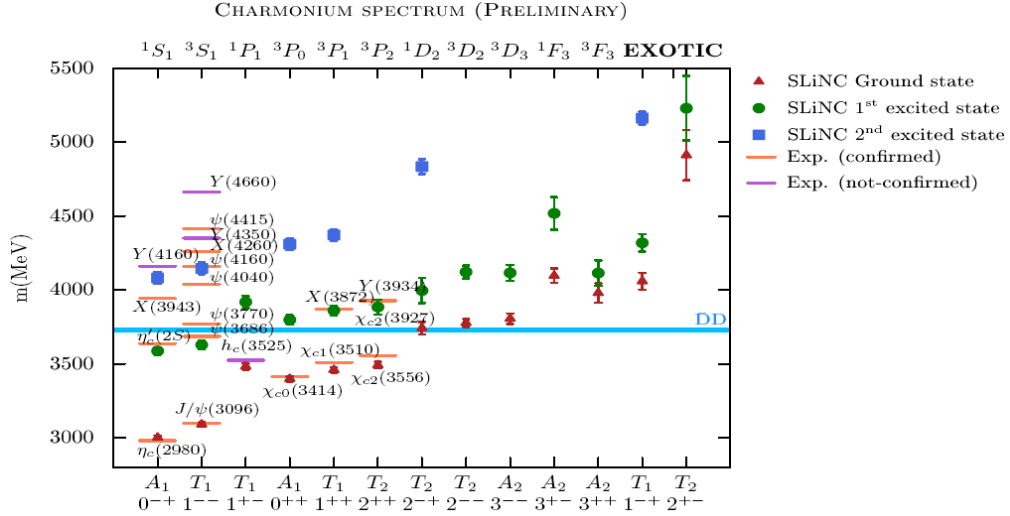}
\caption{Preliminary results for the Charmonium spectrum calculated on SLiNC configurations \cite{Cundy:2009yy} generated by the QCDSF collaboration. Figure taken from \cite{Bali:2011dc}}
\label{fig:spectrum_charm_BMW_QCDSF}
\end{figure}

Figure \ref{fig:spectrum_charm_BMW_QCDSF} shows preliminary results for the charmonium spectrum calculated on QCDSF configurations compared to the experimental spectrum (with a particular assignment for some controversial states). A large number of states including several excited states were extracted and the results agree qualitatively with experiment for states below the $\bar{D}D$ threshold.

\begin{figure}[htb]
\centering
\includegraphics[height=70mm,clip]{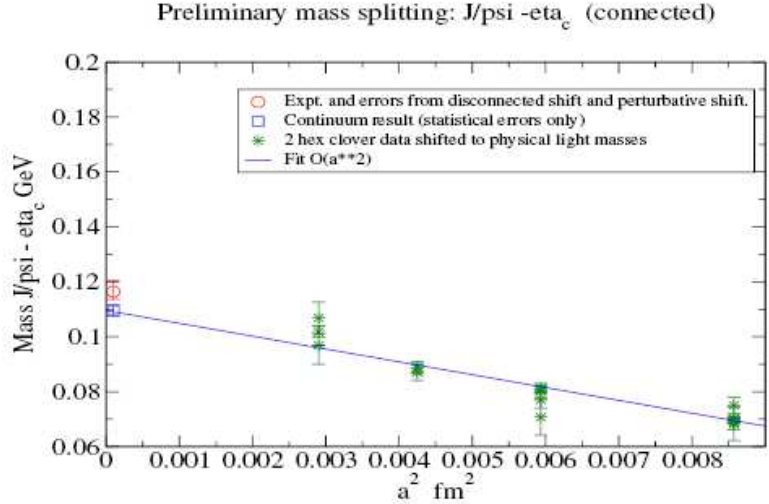}
\caption{Preliminary results for the charmonium hyperfine splitting calculated on the HEX-smeared Wilson-Clover gauge configurations generated by the BMW collaboration. Figure taken from \cite{Bali:2011dc}}
\label{fig:hf_BMW_QCDSF}
\end{figure}

Figure \ref{fig:hf_BMW_QCDSF} shows preliminary results for the charmonium hyperfine splitting calculated on the HEX-smeared Wilson-Clover gauge configurations generated by the BMW collaboration. The results show a strong dependence on the lattice spacing for this action. This is in stark contrast to the results employing improved heavy-quark actions. Nevertheless a value compatible with the experimental hyperfine splitting can be obtained when performing the proper continuum extrapolation


\section{\label{conclusion}Concluding remarks}

In recent years there has been renewed interest in lattice calculations of the charmonium and heavy-light meson spectrum. The availability of configurations with physical or close-to-physical sea quarks, improved heavy quark actions and new calculational techniques like distillation promise many improvements to existing results in the near future. While the era of precision calculations using Lattice QCD has barely begun, there already is remarkable success. In particular, calculations at or around the physical pion mass are becoming common place.  There are multiple collaborations generating gauge configurations with 2+1 or 2+1+1 dynamical flavors suitable for charmonium physics. This rapid progress is also highlighted by several more recent charmonium results \cite{Becirevic:2012dc,Briceno:2012wt,Donald:2012ga} which have appeared since the talk covered in these proceedings was delivered. 

Furthermore for states below inelastic thresholds, the extraction of resonance properties \cite{Luscher:1985dn,Luscher:1986pf,Luscher:1990ux,Luscher:1991cf} from the lattice is within reach. In the charmonium spectrum this necessitates the inclusion of DD, DD* and other states and a first step in this direction has already been taken \cite{Bali:2011rd}. Nevertheless there is still much work ahead with regard to QCD calculations of hadronic excitations.

\Acknowledgements
I would like to thank C. DeTar, Sasa Prelovsek and Richard Woloshyn for helpful discussions. This work is supported in part by the Natural Sciences and Engineering Research Council of Canada (NSERC).


\newpage

\begin{thebibliography}{99}


\bibitem{Godfrey:1985xj} 
  S.~Godfrey and N.~Isgur,
  Phys.\ Rev.\ D {\bf 32}, 189 (1985).


\bibitem{Aoki:2009ix} 
  S.~Aoki {\it et al.}  [PACS-CS Collaboration],
  Phys.\ Rev.\ D {\bf 81}, 074503 (2010)
  [arXiv:0911.2561 [hep-lat]].

\bibitem{Durr:2010vn} 
  S.~D\"urr, Z.~Fodor, C.~Hoelbling, S.~D.~Katz, S.~Krieg, T.~Kurth, L.~Lellouch and T.~Lippert {\it et al.},
  Phys.\ Lett.\ B {\bf 701}, 265 (2011)
  [arXiv:1011.2403 [hep-lat]].


\bibitem{Aoki:2008sm} 
  S.~Aoki {\it et al.}  [PACS-CS Collaboration],
  Phys.\ Rev.\ D {\bf 79}, 034503 (2009)
  [arXiv:0807.1661 [hep-lat]].

\bibitem{Durr:2008zz} 
  S.~D\"urr, Z.~Fodor, J.~Frison, C.~Hoelbling, R.~Hoffmann, S.~D.~Katz, S.~Krieg and T.~Kurth {\it et al.},
  Science {\bf 322}, 1224 (2008)
  [arXiv:0906.3599 [hep-lat]].


\bibitem{Luscher:1985dn} 
  M.~L\"uscher,
  Commun.\ Math.\ Phys.\  {\bf 104}, 177 (1986).

\bibitem{Gattringer:2010zz} 
  C.~Gattringer and C.~B.~Lang,
  Lect.\ Notes Phys.\  {\bf 788}, 1 (2010).


\bibitem{Luscher:1990ck} 
  M.~L\"uscher and U.~Wolff,
  Nucl.\ Phys.\ B {\bf 339}, 222 (1990).


\bibitem{Michael:1985ne} 
  C.~Michael,
  Nucl.\ Phys.\ B {\bf 259}, 58 (1985).

\bibitem{Blossier:2009kd} 
  B.~Blossier, M.~Della Morte, G.~von Hippel, T.~Mendes and R.~Sommer,
  JHEP {\bf 0904}, 094 (2009)
  [arXiv:0902.1265 [hep-lat]].

\bibitem{Burch:2009az} 
  T.~Burch, C.~DeTar, M.~Di Pierro, A.~X.~El-Khadra, E.~D.~Freeland, S.~Gottlieb, A.~S.~Kronfeld and L.~Levkova {\it et al.},
  Phys.\ Rev.\ D {\bf 81}, 034508 (2010)
  [arXiv:0912.2701 [hep-lat]].

\bibitem{Aubin:2004wf} 
  C.~Aubin, C.~Bernard, C.~DeTar, J.~Osborn, S.~Gottlieb, E.~B.~Gregory, D.~Toussaint and U.~M.~Heller {\it et al.},
  Phys.\ Rev.\ D {\bf 70}, 094505 (2004)
  [hep-lat/0402030].

\bibitem{Bernard:2001av} 
  C.~W.~Bernard, T.~Burch, K.~Orginos, D.~Toussaint, T.~A.~DeGrand, C.~E.~Detar, S.~Datta and S.~A.~Gottlieb {\it et al.},
  Phys.\ Rev.\ D {\bf 64}, 054506 (2001)
  [hep-lat/0104002].

\bibitem{Beringer:1900zz} 
  J.~Beringer {\it et al.}  [Particle Data Group Collaboration],
  Phys.\ Rev.\ D {\bf 86}, 010001 (2012).

\bibitem{BESIII:2011ab} 
  M.~Ablikim {\it et al.}  [BESIII Collaboration],
  Phys.\ Rev.\ Lett.\  {\bf 108}, 222002 (2012)
  [arXiv:1111.0398 [hep-ex]].

\bibitem{ElKhadra:1996mp} 
  A.~X.~El-Khadra, A.~S.~Kronfeld and P.~B.~Mackenzie,
  Phys.\ Rev.\ D {\bf 55}, 3933 (1997)
  [hep-lat/9604004].

\bibitem{Follana:2010zz} 
  E.~Follana [HPQCD Collaboration],
  PoS LATTICE {\bf 2010}, 305 (2010).

\bibitem{Levkova:2010ft} 
  L.~Levkova and C.~DeTar,
  Phys.\ Rev.\ D {\bf 83}, 074504 (2011)
  [arXiv:1012.1837 [hep-lat]].

\bibitem{Peardon:2009gh} 
  M.~Peardon {\it et al.}  [Hadron Spectrum Collaboration],
  Phys.\ Rev.\ D {\bf 80}, 054506 (2009)
  [arXiv:0905.2160 [hep-lat]].

\bibitem{Liu:2012ze} 
  L.~Liu {\it et al.}  [for the Hadron Spectrum Collaboration],
  JHEP {\bf 1207}, 126 (2012)
  [arXiv:1204.5425 [hep-ph]].

\bibitem{Mohler:2011ke} 
  D.~Mohler and R.~M.~Woloshyn,
  Phys.\ Rev.\ D {\bf 84}, 054505 (2011)
  [arXiv:1103.5506 [hep-lat]].

\bibitem{Mohler:2012na} 
  D.~Mohler, S.~Prelovsek and R.~M.~Woloshyn,
  arXiv:1208.4059 [hep-lat].

\bibitem{Bernard:2010fr} 
  C.~Bernard {\it et al.}  [Fermilab Lattice and MILC Collaborations],
  Phys.\ Rev.\ D {\bf 83}, 034503 (2011)
  [arXiv:1003.1937 [hep-lat]].

\bibitem{Hasenfratz:2007rf} 
  A.~Hasenfratz, R.~Hoffmann and S.~Schaefer,
  JHEP {\bf 0705}, 029 (2007)
  [hep-lat/0702028].

\bibitem{Hasenfratz:2008fg} 
  A.~Hasenfratz, R.~Hoffmann and S.~Schaefer,
  Phys.\ Rev.\ D {\bf 78}, 014515 (2008)
  [arXiv:0805.2369 [hep-lat]].

\bibitem{Hasenfratz:2008ce} 
  A.~Hasenfratz, R.~Hoffmann and S.~Schaefer,
  Phys.\ Rev.\ D {\bf 78}, 054511 (2008)
  [arXiv:0806.4586 [hep-lat]].

\bibitem{Olsen:2012xn} 
  S.~L.~Olsen [BESIII Collaboration],
  arXiv:1203.4297 [nucl-ex].

\bibitem{Follana:2006rc} 
  E.~Follana {\it et al.}  [HPQCD and UKQCD Collaborations],
  Phys.\ Rev.\ D {\bf 75}, 054502 (2007)
  [hep-lat/0610092].

\bibitem{Dudek:2010wm} 
  J.~J.~Dudek, R.~G.~Edwards, M.~J.~Peardon, D.~G.~Richards and C.~E.~Thomas,
  Phys.\ Rev.\ D {\bf 82}, 034508 (2010)
  [arXiv:1004.4930 [hep-ph]].

\bibitem{Bali:2011dc} 
  G.~Bali, S.~Collins, S.~D\"urr, Z.~Fodor, R.~Horsley, C.~Hoelbling, S.~D.~Katz and I.~Kanamori {\it et al.},
  PoS LATTICE {\bf 2011}, 135 (2011)
  [arXiv:1108.6147 [hep-lat]].

\bibitem{Durr:2010aw} 
  S.~D\"urr, Z.~Fodor, C.~Hoelbling, S.~D.~Katz, S.~Krieg, T.~Kurth, L.~Lellouch and T.~Lippert {\it et al.},
  JHEP {\bf 1108}, 148 (2011)
  [arXiv:1011.2711 [hep-lat]].

\bibitem{Cundy:2009yy} 
  N.~Cundy, M.~Gockeler, R.~Horsley, T.~Kaltenbrunner, A.~D.~Kennedy, Y.~Nakamura, H.~Perlt and D.~Pleiter {\it et al.},
  Phys.\ Rev.\ D {\bf 79}, 094507 (2009)
  [arXiv:0901.3302 [hep-lat]].

\bibitem{Becirevic:2012dc} 
  D.~Becirevic and F.~Sanfilippo,
  arXiv:1206.1445 [hep-lat].

\bibitem{Briceno:2012wt} 
  R.~A.~Briceno, H.~-W.~Lin and D.~R.~Bolton,
  arXiv:1207.3536 [hep-lat].

\bibitem{Donald:2012ga} 
  G.~C.~Donald, C.~T.~H.~Davies, R.~J.~Dowdall, E.~Follana, K.~Hornbostel, J.~Koponen, G.~P.~Lepage and C.~McNeile,
  arXiv:1208.2855 [hep-lat].

\bibitem{Luscher:1985dn} 
  M.~L\"uscher,
  Commun.\ Math.\ Phys.\  {\bf 104}, 177 (1986).


\bibitem{Luscher:1986pf} 
  M.~L\"uscher,
  Commun.\ Math.\ Phys.\  {\bf 105}, 153 (1986).

\bibitem{Luscher:1990ux} 
  M.~L\"uscher,
  Nucl.\ Phys.\ B {\bf 354}, 531 (1991).

\bibitem{Luscher:1991cf} 
  M.~L\"uscher,
  Nucl.\ Phys.\ B {\bf 364}, 237 (1991).

\bibitem{Bali:2011rd} 
  G.~S.~Bali, S.~Collins and C.~Ehmann,
  Phys.\ Rev.\ D {\bf 84}, 094506 (2011)
  [arXiv:1110.2381 [hep-lat]].


\end{thebibliography}
\end{document}